\documentclass[letterpaper]{article}
\usepackage{aaai20}
\usepackage{times}
\usepackage{helvet}
\usepackage{courier}
\usepackage[hyphens]{url}
\usepackage{graphicx}
\usepackage{graphicx}
\usepackage{bm}
\usepackage{amsmath}
\usepackage{mathtools}
\usepackage{subcaption}
\usepackage{array,booktabs,ragged2e}

\newcommand{\E}{\mathrm{E}}

\title{Learning Fair and Interpretable Representations via Linear Orthogonalization}

\author{
Yuzi He, \textsuperscript{\rm 1,}\textsuperscript{\rm 2}
Keith Burghardt, \textsuperscript{\rm 1}
Kristina Lerman, \textsuperscript{\rm 1} \\
\textsuperscript{\rm 1} Information Sciences Institute\\University of Southern California\\4676 Admiralty Way, Suite 1001
\\Marina del Rey, CA 90292\\
\textsuperscript{\rm 2} Department of Physics \& Astronomy\\University of Southern California\\Los Angeles, CA 90089-0484\\
yuzihe@usc.edu, keithab@isi.edu, lerman@isi.edu
}

\nocopyright
\begin{document}
\maketitle
\begin{abstract}
To reduce human error and prejudice, many high-stakes decisions have been turned over to machine algorithms. However, recent research suggests that this \emph{does not} remove discrimination, and can perpetuate harmful stereotypes. While algorithms have been developed to improve fairness, they typically face at least one of three shortcomings: they are not interpretable, their prediction quality deteriorates quickly compared to unbiased equivalents, and 
they are not easily transferable across models
. To address these shortcomings, we propose a geometric method that removes correlations between data and any number of protected variables. Further, we can control the strength of debiasing through an adjustable parameter to address the trade-off between prediction quality and fairness. The resulting features are interpretable and can be used with many popular models, such as linear regression, random forest, and multilayer perceptrons. The resulting predictions are found to be more accurate and fair compared to several state-of-the-art fair AI algorithms across a variety of benchmark datasets. Our work shows that debiasing data is a simple and effective solution toward improving fairness.
\end{abstract}

\section{Introduction}
Machine learning (ML) models sift through mountains of data to make decisions on matters big and small: e.g., who should be shown a product, hired for a job, or given a home loan. Machine inference can systematize decision processes to take into account orders of magnitude more information, produce accurate decisions, and avoid the common pitfalls of human judgment, such as belief in a just world or selective attention \cite{Kruglanski1983}. 
Moreover, unlike people, machines will never make poor decisions when tired~\cite{Danziger2011}, pressed for time or distracted by other matters~\cite{shah2012some,mani2013poverty}.

Recent research suggests, however, that discrimination remains pervasive~\cite{angwin2016machine,chouldechova2017fair,dressel2018accuracy,mathdestruction}: for example, a model used to evaluate criminal defendants for recidivism assigned systematically higher risk scores to African Americans than to Caucasians \cite{angwin2016machine}. As a result, reformed African American defendants, who would never commit another crime, were deemed by the model to present a higher risk to society---as much as twice as high~\cite{angwin2016machine,dressel2018accuracy}---as reformed white defendants, with potentially grave consequences on how they were treated by the justice system. 

The emerging field of AI fairness has suggested ways to mitigate harmful model biases~\cite{dwork2012fairness,chouldechova2017fair,alex2018frontiers}, e.g., penalizing unfair inferences~\cite{dwork2012fairness,Berk2017}, or creating representations    that do not strongly depend on protected features~\cite{Jaiswal2018,Moyer2018,locatello2019fairness}. These methods, however, fall short in one or more critical dimensions: interpretability, prediction quality, and generalizability. 
We define \emph{interpretability} as the ability to understand how features affect---or bias---a model's predicted outcome. Interpretability is needed to 
improve transparency and accountability of AI systems. While models must sacrifice \emph{prediction quality} (as  measured by accuracy, mean squared error, or another metric) to improve fairness~\cite{Pierson2017}, the trade-off does not need to be as drastic as what current methods achieve. Finally, we define \emph{generalizability} as the ability to easily apply fairness algorithms across multiple models and datasets. In contrast, state-of-the-art fairness methods  are specialized to linear regressions or random forests \cite{Zafar2017,Kamiran2010,Berk2017}. Similarly,  methods that create fair latent features for neural networks (NN) \cite{Jaiswal2018,Moyer2018} cannot be easily applied to improve fairness in non-NN models. These fair AI algorithms were not meant to be generalizable because there does not seem to be adequate meta-algorithms that debias a whole host of ML models. One might naively expect that we can just create a single fair model and apply it to all datasets. The problem is that model performance varies greatly on different datasets. While NNs are critical for, e.g., image recognition \cite{Ciregan2012}, other methods perform better for small data \cite{Olson2018}, especially when the number of dimensions is high and the sample size low~\cite{Liu2017}. There is no one-size-fits-all model and there is no one-size-fits-all model debiasing method. Is there an easier way to create fairer predictions other than specialized methods for specialized ML models? Chen et al. offer some clues to addressing this fundamental issue in fair AI \cite{Chen2018}: by addressing data biases, we can potentially improve fair AI across the spectrum of models, and achieve fairness without greatly sacrificing  prediction quality.

Inspired by these ideas, we describe a geometric method  
for \textit{debiasing features}. Depending on the hyperparameter we choose, these features are mathematically guaranteed to be 
uncorrelated with specified sensitive, or \textit{protected}, features. This method is exceedingly fast and the debiased features are highly correlated with the original features (average Pearson correlations are between 0.993--0.994 across the three datasets studied in this paper). These debiased features are as interpretable as the original features when applied to any model. When applied to linear regression, for example, the coefficients are the same or similar to the coefficients of the original features when controlling for protected variables (see Methods). 
These debiased features serve as a fair representation of data that can be used with a number of NN and non-NN ML models, such as linear regression, random forest, support vector machines (SVMs), and multilayer perceptrons (MLPs). 
While previous methods have created fair representations \cite{Olfat2018,Samadi2018,Jaiswal2018,Moyer2018}, these methods create representations that are either not very interpretable, like PCA components, or the relationship between these fair representations and the original features have not been established. 
We evaluate the proposed approach on several benchmark datasets. We show that models using these debiased features are more accurate for almost any level of fairness we desire.

In the rest of the paper, we first review recent advances in fair AI to highlight the novelty of our method. Next, we describe in the Methods section our methodology to improve data fairness, and the definitions of fairness we use in the paper. In  Results, we describe how our method improves fairness in both synthetic data and empirical benchmark data. We compare to several competing methods and demonstrate the advantages of our method. Finally, we summarize our results and discuss future work in the Conclusion section.


\section{Related Work}


Social scientists use linear regression for data analysis due to its simplicity and interpretability. 
Interpretability comes from regression coefficients, which specify how the outcome, or response, changes when features change by one unit. 
However, regression creates unfair outcomes, even when protected features are excluded from the model, because other features may be correlated with them.

To make regression models fair, researchers introduced a loss function to penalize regression for unfair outcomes~\cite{Berk2017}. Similarly, \cite{Zafar2015} created fair logistic regression by introducing fairness constraints that limit the covariance between protected features and the outcome. An alternate method achieved fairness by constraining false positive or false negative rates~\cite{Zafar2016}. There are some issues in these works, however. First, protected features are not included in the logistic model with fairness constraints. While this improves privacy, it forces the parameters of logistic models to take certain combinations which will minimize the correlation with the protected features. This can reduce the accuracy when the constraints are strict. The issue for the second method is mainly numeric. The algorithm requires an optimization of a convex loss function on a non-convex parameter space. While these models are generally interpretable, the approaches do not transfer to other models. 

Researchers have explored a variety of fair data representation methods~\cite{Jaiswal2018,Moyer2018,Louizos2015,Xie2017,Zemel2013,Samadi2018,Olfat2018}. Some of those works use NNs to embed raw features in a lower-dimensional space, such that the embedding will contain the information about the outcome variable, but at the same time, contain little information about the protected feature. Fair logistic models or fair scoring, on the other hand, can be regarded as a one dimensional embedding of data, which makes sure that the predictions, $\hat y$, are independent of the protected features. They are mainly used with NNs, which are accurate but often lack interpretability. Two methods were instead developed to improve fairness of PCA features \cite{Samadi2018,Olfat2018}. While they can be applied to many ML models, they lack interpretability compared to the original features.


\citeauthor{Johndrow2017} (\citeyear{Johndrow2017}) proposed an algorithm which removes sensitive information about protected groups based on inverse transform sampling. The algorithm transforms individual features such that the transformed features satisfy the marginal distribution. Although this method can guarantee that predictions are fair in a probabilistic sense, it has a critical disadvantage --- as the number of protected features $n_p$ increases, the number of protected groups increases as $O(2^{n_p})$. This means that in order to properly estimate conditional and marginal distribution of features, one needs exponentially increasing population size. Our method overcomes these difficulties by using linear algebra as the basis for learning unbiased representations. This allows our algorithm to only take $O(n_p^2)$ time to debias data. Moreover, our method is a white box: it is  interpretable and can be fully scrutinized, unlike a black box method.

\section{Methods} 
We describe a geometric method for constructing fair interpretable representations. These representations can be used with a variety of ML methods to create fairer accurate models of data.

\subsection{Fair Interpretable Representations}
We consider tabular data with $n$ entries and $m$ features. The features are vectors in the $n$-dimensional space, denoted as $\bm x_i$ where $i = 1, 2, \cdots, m$, and one of the columns corresponds to the outcome, or target variable $\bm y$. Among the features, there are also $n_p$ protected features, $\bm p_i, i = 1,\dots,n_p$. As a pre-processing step, all features are centered around the mean: $\langle x_i\rangle = 0$.

We describe a procedure to debias the data so as to create linearly fair features. 
We aim to construct a representation $\bm{r}_j$ of a feature $\bm x_j$, that is uncorrelated with $n_p$ protected columns $\bm p_i, i = 1, \dots n_p$, but highly correlated to feature $\bm x_j$. We recall that Pearson correlation between the representation $r_j$ and any feature $x_k$ is defined as $$\text{Corr}(r_j,x_k) = (\E[r_j \cdot x_k]-\E[r_j]\E[x_k])/(\sigma_{r_j}\sigma_{x_k}),$$ where $\E[.]$ is the expectation, and $\sigma_{r_j} = \sqrt{\E[r_j^2]-\E[r_j]^2}$ and $\sigma_{x_k} = \sqrt{\E[x_k^2]-\E[x_k]^2}.$ Because all the features are centered (and we also assume that $r_j$ is centered), $\E[r_j]=\E[x_k]=0$, we have $$\sigma_{r_j} = \sqrt{\E[r_j^2]} = \| r_j \|/\sqrt n,$$ $$\sigma_{x_k} = \sqrt{\E[x_k^2]} = \| x_k \|/\sqrt n$$ and $$\E[r_j \cdot x_k]=\bm{r}_j\cdot\bm{x}_k/n.$$ Therefore $$\text{Corr}(r_j,p_i) = \bm{r}_j\cdot\bm{p}_i/(\|r_j\| \cdot \| p_i \|)$$ and $$\text{Corr}(r_j,x_j) = \bm{r}_j\cdot\bm{x}_j /(\|r_j\| \cdot \| x_j \|).$$ Zero correlations between $r_j$ and $n_p$ protected columns requires that $r_j$ lives in the solution space of $\bm{r}_j \cdot \bm{p}_i = 0, i = 1 \dots n_p$. Maximizing correlations between $r_j$ and $x_j$ under this constraint is equivalent to projecting $\bm x_j$ into the solution space of $\bm{r}_j \cdot \bm{p}_i = 0, i = 1 \dots n_p$. 

To calculate $\bm r_j$, we can first create an orthonormal basis of vectors $\bm p_i$, which we can label as $\bm {\bar p}_i$. We then construct a projector $P_f = \sum_{i=1}^{n_p} \bm{\bar p}_i \bm{\bar p}_i^{T}$. The representation $\bm r$ is given as 
\begin{equation} 
  \bm r_j = \bm x_j- P_f \bm x_j = (I - P_f) \bm x_j.
  \label{eq:mt_5}
\end{equation}
Using the Gram–Schmidt process, the orthonormal basis can be constructed in $O(n\times n_p^2)$ time and for every fair representation of features, the projection takes $O(n\times n_p)$ time. Given $n_f$ features, the total time of the algorithm is $O(n\times n_f\times n_p^2)$ Therefore our method scales linearly with respect to the size of the data and the number of features. In practice, this is exceedingly fast. For example, this algorithm only takes less than 200 milliseconds to run on the Adult dataset described below, which has 45K rows, 103 unprotected features, and 1 protected feature.

While the previous discussion was on how to create linearly fair features, one can make linearly fair outcome variables, $\bm r_y$ through the same process. In prediction tasks, however, we do not have access to the outcome data. While our method does not guarantee that every model's estimate of the outcome variable, $\hat y$ is fair, we find that it can significantly improve the fairness compared to competing methods. Moreover, in the special case of linear regression, it can be shown that the resulting estimate, $\hat y$, is uncorrelated with the protected variables.

Inevitably, the prediction quality of a model using such linearly fair features will drop compared to using the original features, because the solution is more constrained. 
To address this issue, we introduce a parameter $\lambda \in [0,1]$, which indicates the fairness level. We define the parameterized latent variable as
\begin{equation}
  \bm r'_j(\lambda) = \bm r_j + \lambda\cdot(\bm x_j - \bm r_j) .
  \label{eq:mt_6}
\end{equation}
Here, $\lambda = 0$ corresponds to $\bm r'_j(\lambda) = \bm r_j$, which is strictly orthogonal to the protected features $\bm p_i$; while $\lambda = 1$ gives $\bm r'(\lambda) = \bm x_j$.

The protected features can be both real valued and cardinal. The fair representation 
method can also handle categorical protected features by introducing dummy variables. Specifically, if a variable $X$ has $k$ categories $x_1,~x_2,...,x_{k}$, we can can convert them to $k-1$ binary variables where the $i^{th}$ variable is 1 if the variable is category $x_i$, and otherwise 0. If all variables are 0, then the category is $x_k$. As a simple example, if a feature $X$ has 3 categories, $x_1$, $x_2$, and $x_3$, then the dummy variables would be $\tilde{x}_1$ and $\tilde{x}_2$. If $\tilde{x}_1=1$, the category is $x_1$, if $\tilde{x}_2=1$, then the category is $x_2$, and otherwise is $x_3$. The condition of fairness in this case is interpreted as same mean value of the latent variables in different categorical groups.

\subsection{Fair Models} 
Using the procedure described above, we can construct a fair representation of every feature, and use the fair features to model the outcome variable. Consider a linear regression model that  includes all  features: $n_p$ protected features $p_i, i=1,\cdots,n_p$ and $n_f=m-n_p$ non-protected features features $x_i, i=1,\dots,n_f$.
\begin{equation}
    \hat y = \beta_0 + \sum_{i=1}^{n_f} \beta_i x_i 
                     + \sum_{i=1}^{n_p} \gamma_i p_i.
\end{equation}
After transforming the features to fair features $x_i'$, the fair regression model reduces to:
\begin{equation}
    \hat y' = \beta_0' + \sum_{i=1}^{n_f} \beta_i' r_i.
\end{equation}
Here, $r_i$ corresponds to the fair versions of $x_i$. We can prove that $\beta_i = \beta_i', i=1,\dots,n_f$, but the predicted value $\hat y'$ is uncorrelated with protected features $p_i, i = 1,\dots,n_p$. In general linear regression, such as logistic regression, this proof does not hold, but we numerically find that coefficients are similar. 

We should take a step back at this point. The fair latent features are close approximations of the original features, therefore we expect that, and in certain cases can prove, that the regression coefficients of the fair features should be approximately the coefficients of the original features. The fair features can, by this definition, be considered almost as interpretable as the original features. 

In addition to regression, fair representations could be used with other ML models, such as AdaBoost \cite{freund1997decision}, NuSVM \cite{chang2011libsvm}, random forest \cite{breiman2001random}, and multilayer perceptrons \cite{Rosenblatt1961}. 

\subsection{Measuring Fairness}
While there exists no consensus for measuring fairness, researchers  have proposed a variety of metrics, some focusing on representations and some on the predicted outcomes~\cite{verma2018fairness,hutchinson201950}. We will therefore compare our method to competing methods using the following metrics: Pearson correlation, mutual information, discrimination, calibration, balance of classes, and accuracy of the inferred protected features. Due to space limitations, we leave mutual information out of our analysis in this paper, and do not compare calibration and balance of classes to model accuracy. Results in all cases are similar.

\subsubsection{Fairness of Outcomes}
One can argue that outcomes are fair if they do not depend on the protected features. If this is the case, a malicious adversary won't be able to guess the protected features from the model's predictions. One way to quantify the dependence is through \textit{Pearson correlation} between (real valued or cardinal) predictions and protected features. For models making binary predictions, fairness can be measured using the \textit{mutual information} between predictions and the protected features, given that protected features are discrete. We find mutual information and Pearson correlations create qualitatively similar findings, despite mutual information being a non-linear metric, therefore we focus on Pearson correlations in this paper. Previous work \cite{Zemel2013} has also defined a \textit{discrimination metric} for binary predictions as below. Consider a protected variable $p_1$, a binary prediction $\hat y$ of an outcome $y$. The metric measures the bias of a binary prediction $\hat y$ with respect to a single binary protected feature $p_1$ using the difference of positive rates between the two groups.
\begin{equation}
    y_\text{Discrim} = \left| 
      \frac{\sum_{n:p_1[n]=0} \hat y[n]}
      {\sum_{n:p_1[n]=0} 1} 
    - \frac{\sum_{n:p_1[n]=1} \hat y[n]} {\sum_{n:p_1[n]=1} 1} \right|
\end{equation}

For real valued predictions ($\hat y \in [0,1]$), \citeauthor{Kleinberg2016}~(\citeyear{Kleinberg2016}) suggested a more nuanced way to measure fairness: 
\begin{itemize}
    \item \textbf{Calibration within groups}: Individuals assigned predicted probability $\hat y \in [r_0 - \delta, r_0 + \delta]$, $(\delta > 0~\text{and}~\delta \ll 1)$ should have an approximate positive rate of $r$. This should hold for both protected groups ($p_1 = 0$ and $p_1 = 1$). 
    \item \textbf{Balance for the negative class}: The mean $\hat y$ of group $p_1 = 0, y = 0$ and group $p_1 = 1, y = 0$ should be the same.
    \item \textbf{Balance for the positive class}: The mean $\hat y$ of group $p_1 = 0, y = 1$ and group $p_1 = 1, y = 1$ should be the same.
\end{itemize}
In some cases, calibration error is difficult to calculate, as it depends on how predictions are binned. In these cases, we can measure calibration error using log-likelihood of the labels given the real valued predictions as a proxy. By definition, logistic regression maximizes the (log-)likelihood function, assuming the observations are sampled from independent Bernoulli distributions where $P(y[n]|X[n]) = \hat y_i[n]$. Better log-likelihood implies that the individuals assigned probabilities $\hat y \in [r_0-\delta, r_0+\delta]$ are more likely to have a positive rate $r$, which is better calibrated according to \citeauthor{Kleinberg2016}

\subsubsection{Fairness of Representations}
Several past studies examined the fairness of representations, arguing that models using fair representations will also make fair predictions. Learned representations are considered fair if they do not reveal any information about the protected features~\cite{Jaiswal2018,Moyer2018,Louizos2015,Xie2017,verma2018fairness}. The studies trained a discriminator to predict protected features from the learned representations---using accuracy as a measure of fairness. 

Following this approach, we treat the predicted probabilities as a one-dimensional representation of data and use the \textit{accuracy of the inferred protected features} as a measure of fairness. However, this method is not effective in situations where the protected classes are unbalanced. Let us assume the fair representation is $R$ and the protected feature is $p_1$. For simplicity, we only consider the case of a single binary protected feature. The discriminator infers the protected feature in a Bayesian way, namely, 
\begin{equation}
    P(p_1=c|R) = \frac{P(R|p_1=c)P(p_1=c)}{P(R)}, c = 0|1 
    \label{eq:method_bayes}
\end{equation}
In the case where there is a large difference between $P(p_1=0)$ and $P(p_1=1)$, even if there is useful information in the distribution $P(R|p_1=c)$, the discriminator will not perform significantly better than the baseline model, the majority class classifier. 

\section{Results}


\subsection{Synthetic Data}
\begin{figure}[!ht]
\centering
    \includegraphics[width=1.0\columnwidth]{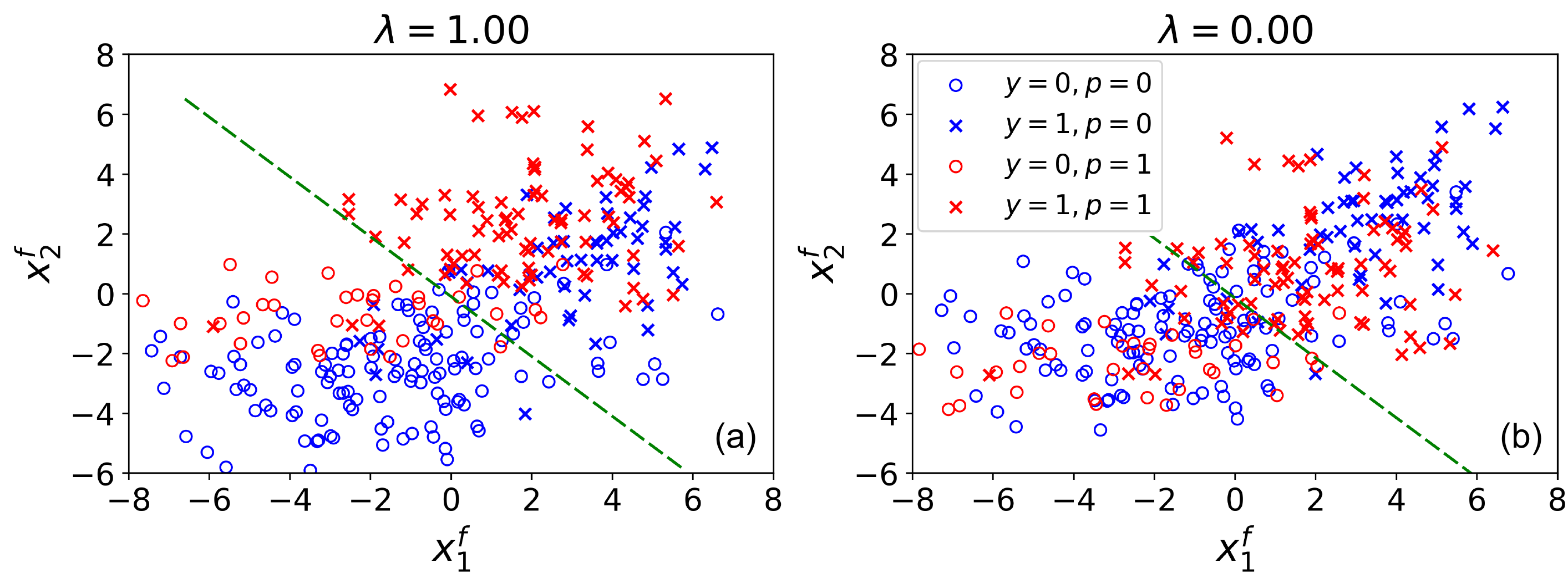}
    \caption{Fair synthetic data. (a) raw data ($\lambda=1.0$), (b) plot for fairness level $\lambda = 0.0$. The two features in the data are $x^f_1$ and $x^f_2$, and the two classes we want to protect are in red and blue. The two outcome classes are represented as two symbols: $\times$ and $\circ$. 
    }
    \label{fig:syn_twox_plot}
\end{figure}

We create synthetic biased data using the procedure described in \cite{Zafar2016}. We generate data with one binary protected variable $s$, one binary outcome $y$, and two continuous features, $x_1$ and $x_2$, which are bivariate Gaussian distributed within each value of $s$. In the Fig.~\ref{fig:syn_twox_plot}, we use color to represent protected feature values (red, blue) and outcome using symbol ($\times$, $\circ$). The first observation is that there is an imbalance in the joint distribution of the protected features and the outcome variable. For blue color markers, there are more blue $\circ$s than blue $\times$s.
We expect that a logistic classifier trained on this data will show similar unbalanced behavior. To demonstrate our method, we choose two different fairness levels, $\lambda=\{0.0,1.0\}$. We first transform the two features into their corresponding fair representations and then we train logistic classifiers using these fair representations. In Fig.~\ref{fig:syn_twox_plot}, we plot the data using the fair representations and we show the classification boundary using a green dashed line. We can observe that for $\lambda=0$, the blue markers and red markers are mixed (less discrimination and bias), but for $\lambda=1.0$ (equivalent to raw data), the blue and red markers tend to separate from each other. We can estimate this imbalance by comparing the ratio of blue in individuals predicted as $\circ$ and the ratio of blue in individuals predicted as $\times$. The larger the difference, the more the imbalance. Quantitatively, for $\lambda = 0.0$, there are 62.7\% blue in o-predictions and 52.9\% in x-predictions. For $\lambda = 1.0$, those ratios are 76.2\% and 36.5\%. The accuracy of outcome predictions are 0.811 and 0.870 for the fair and original features, respectively, thus demonstrating that, while increasing fairness does indeed sacrifice in accuracy, the loss can be relatively small. 
Overall, the results suggest that biased data creates biased models, but our method can make fairer models.

We demonstrate how our method can achieve fair classification using synthetic data (see Appendix), and also compare our prediction quality and fairness to other fair AI algorithms using benchmark datasets.

\subsection{Real-World Data}

\textbf{German} dataset has 61 features about 1,000 individuals, with a binary outcome variable denoting whether an individual has a good credit score or not. The protected feature is gender. (\url{https://archive.ics.uci.edu/ml/datasets/statlog+(german+credit+data)})

\textbf{COMPAS} dataset contains data about 6,172 defendants. The binary outcome variable denotes whether the defendant will recidivate (commit a crime) within two years. The protected feature is race (whether the race is African American or not), and there are nine features in total. (\url{https://github.com/propublica/compas-analysis})

\textbf{Adult} dataset contains data about 45,222 individuals. The outcome variable is binary, denoting whether an individual has more than \$50,000. The protected feature is age, and there are 104 features in total. (\url{https://archive.ics.uci.edu/ml/datasets/Adult})

Debiased features had mean correlations of 0.993, 0.994, and 0.994, for the German, COMPAS, and Adult data, respectively. 
We reserved $20\%$ of the data in the Adult and COMPAS datasets for testing and used the remaining data to perform 5-fold cross validation. This ensured no leakage of information from the training set to the testing set. 
 The German dataset is much smaller than the rest, so it was randomly divided into five folds of training, validation and testing sets. Each set had 50\%, 20\% and 30\% of all the data. We measured the performance metrics on the test data.

We varied the fairness parameter $\lambda$ between 0 and 1 and applied the debiased features to logistic regression, AdaBoost, 
NuSVM, 
random forest, and multilayer perceptrons. 
In practice, one could use a host of commercial ML models and pick the most accurate one given their fairness tolerance.

\begin{figure*}[!h]
\centering
        \includegraphics[width=1.0\textwidth]{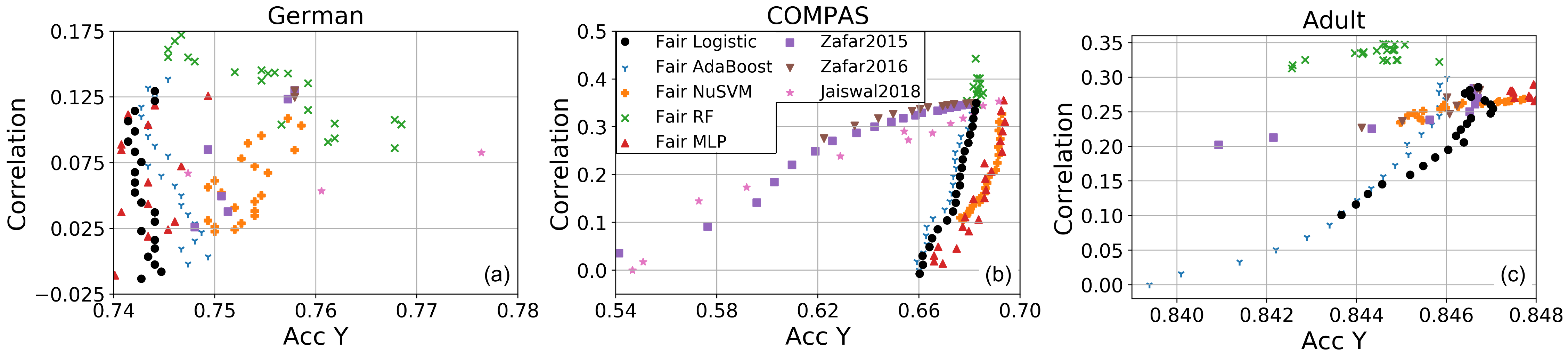}
        \caption{Fairness versus accuracy. Plots show  Pearson correlation versus accuracy of predictions (\textit{Acc Y}) for the German, COMPAS and Adult datasets. For each plot, \textit{Zafar2015} stands for  \cite{Zafar2015}, \textit{Zafar2016} for  \cite{Zafar2016} and \textit{Jaiswal2018} for \cite{Jaiswal2018}. \textit{Fair NuSVM}, \textit{Fair RF}, \textit{Fair AdaBoost}, and \textit{Fair MLP} results are produced using the fair representations constructed by our proposed method with NuSVM~\cite{chang2011libsvm}, random forest~\cite{breiman2001random},  AdaBoost~\cite{freund1997decision}, and multilayer perceptrons \cite{Rosenblatt1961} models, respectively. 
        The results of UAI are not shown for the Adult dataset, since its best accuracy (0.83) lies outside of the boundary of the  plot. (Same for Figure \ref{fig:acc_disc} and \ref{fig:acc_yp}.)}
        \label{fig:acc_corr}
        \end{figure*}
\subsection{Comparison Against State-of-the-Art}
We compared our method to several previous fair AI algorithms. For the models proposed by \cite{Zafar2015,Zafar2016}, we vary the fairness constraints from perfect fairness to unconstrained. For the ``Unified Adversarial Invariance'' (UAI) model proposed by \cite{Jaiswal2018}, we vary the $\delta$ term in the loss function from 0 (no fairness) to very large value, e.g., $9.0 \times 10^{19}$ for COMPAS dataset, (large $\delta$ value corresponds to perfect fairness). The predictions of the UAI model for the German and Adult datasets are provided by the authors.  We are interested in  (1) how different models tradeoff between accuracy and fairness and (2) how different metrics of fairness compare to each other.

\begin{figure*}[!ht]
    \centering
        \includegraphics[width=1.0\textwidth]{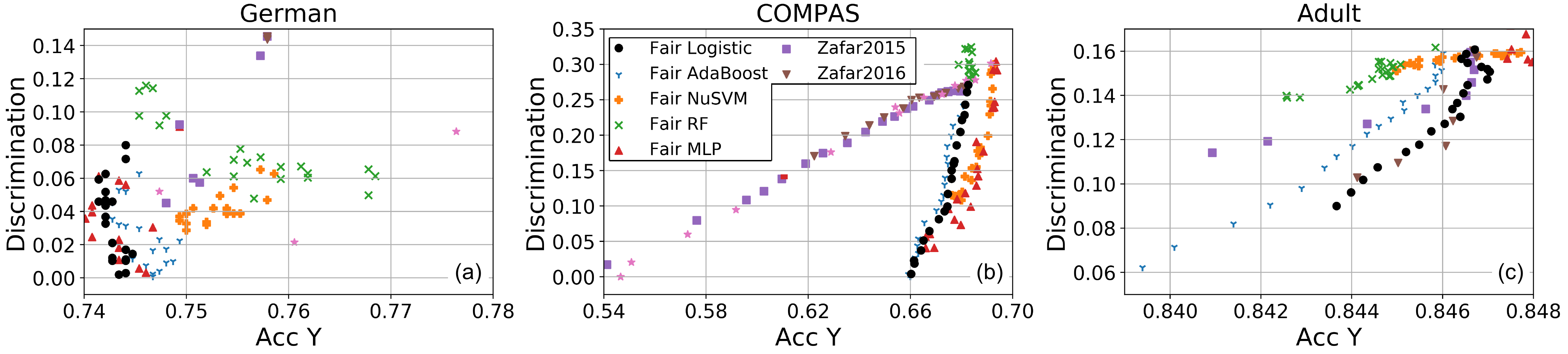}
        \caption{Discrimination versus accuracy plots for the three datasets. 
        }
        \label{fig:acc_disc}
        \end{figure*}

\paragraph{Fairness Versus Accuracy}
We first investigate the trade-offs between prediction accuracy ($Acc\ Y$) and fairness, which we measure three different ways: (1) Pearson correlation between the protected feature and model predictions, 
(2) discrimination between the binary protected feature and the binarized predictions (predicted probabilities above 1/2 are given a value of 1, and are otherwise 0) and (3) the accuracy of predicting protected features from the predictions 
($Acc\ P$). To robustly predict the protected features from the model predictions, 
we used both a NN with three hidden layers, which is used by former works \cite{Jaiswal2018,Moyer2018,Louizos2015,Xie2017,Zemel2013} and a random forest model. We report the better accuracy of those two models. Figure~\ref{fig:acc_corr},\ref{fig:acc_disc} and \ref{fig:acc_yp} shows the resulting comparisons.

The figures show that models using the proposed fair features achieve significantly higher accuracy---for the same degree of fairness---compared to competing methods. Equivalently, we achieve greater fairness with equivalent accuracy. In Fig.~\ref{fig:acc_yp}, we find \textit{Acc P} shows little difference from the baseline majority class classifier for the German and Adult datasets. The reason is explained in Eq.(\ref{eq:method_bayes}). On the other hand, \textit{Acc P} of COMPAS dataset shows a clear trend because the majority baseline is around 0.51, which is consistent with the Eq.(\ref{eq:method_bayes}).  For the Adult dataset, the fair logistic regression cannot achieve perfect fairness but the situation is improved by AdaBoost. We discover, in other words, that there is no single ML model that achieves greater accuracy for a given value of fairness, but our method allows us to choose suitable models to achieve greater accuracy.

\subsubsection{Fairness of Representations}
    \begin{figure*}[!ht]
    \centering
            \includegraphics[width=1.0\textwidth]{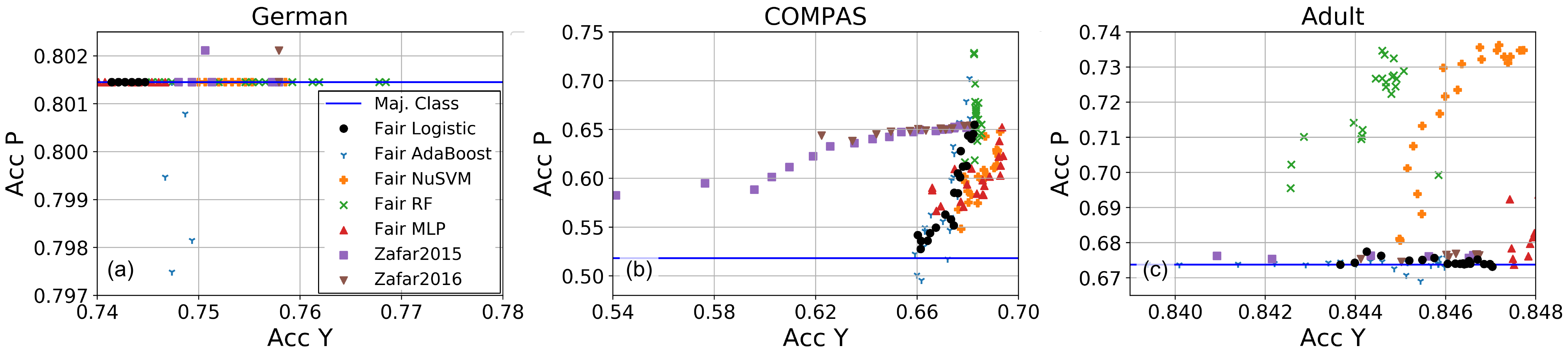}
        \caption{Accuracy of inferring the protected variable from the model's predictions ($Acc~P$) versus the accuracy of predicting the outcome ($Acc~Y$) for the three datasets.
       }
        \label{fig:acc_yp}
        \end{figure*}

We compared our method to earlier works using fair representations. Previous works used NNs to encode the features into a high dimensional embedding space and then separately trained discriminators to infer the protected feature and the outcome variables. The accuracy of inferring protected feature and outcome are reported. Ideally, the accuracy for the outcome should be high and the accuracy of inferring the protected features should be close to the majority class baseline. We set the fairness level to $\lambda = 0$ (perfect fairness). We show \textit{Acc P} and \textit{Acc Y} for various methods in Table~\ref{tb:acc_yp_german_adult} (Appendix) and Fig.~\ref{fig:acc_yp}. Our method applied to a logistic model has similar fairness to the best existing methods but is very fast, easy to understand, and creates more interpretable features.

\subsubsection{Balance Versus Calibration}
\begin{figure*}[!ht]
\centering
    \includegraphics[width=1.0\textwidth]{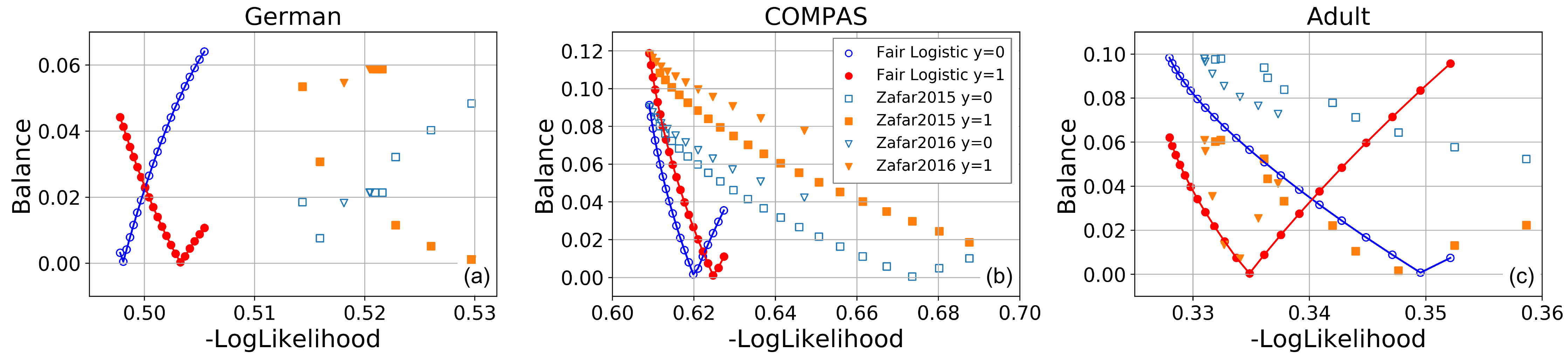}
    \caption{Balance vs. negative log-likelihood (calibration error) for the German, COMPAS and Adult datasets. In the plot, there are two sets of curves for every model, labeled \textit{$y=0$} and \textit{$y=1$}. \textit{$y=0$} stands for the difference of mean $\hat y$ (between different protected classes) given to the individuals with negative $y=0$, and \textit{$y=1$} stands for individuals with positive outcomes $y=1$. (These differences are called balance of negative or positive class by \cite{Kleinberg2016}.) Fairer models are those in the lower left corner of each plot.}
    \label{fig:balance-calibration}
\end{figure*}

Finally, we use another measure of fairness that captures the degree to which each model makes mistakes. Figure~\ref{fig:balance-calibration} shows delta score (i.e., balance) versus negative log-likelihood (i.e., calibration error). Fairer predictions are located in the lower left corner of each figure
, meaning that there are fewer differences in outcomes for the different classes. 
We only compare the logistic model with fair features to the models proposed by Zafar et al.~\cite{Zafar2015,Zafar2016}, because these models maximize the log-likelihood function (minimize calibration error) when selecting parameters. 
For all datasets, our method generally achieves greater fairness.

\section{Conclusion}

We show that our algorithm simultaneously achieves three advances over many previous fair AI algorithms. First, it is interpretable; the features we construct are minimally affected by our fair transform. While this does not mean the models trained on these features are interpretable (they could be a black box), it does mean that any method used to interpret features could easily be used for these fairer features as well. Next, the features better preserve model prediction quality. Namely, models using these features were more accurate than competing methods when the value of the fairness metric was held fixed. This is in part due to the third principle: that our method can be applied to any number of commercial models; it merely acts as a pre-processing step. Different models have different strengths and weaknesses; while some are more accurate, others are fairer. We can pick and choose particular models that achieve both high fairness and accuracy, whether it is a linear model like logistic regression or a non-linear model like a multilayer perceptron, as shown in Figs.~\ref{fig:acc_corr},~\ref{fig:acc_disc}., \&~\ref{fig:acc_yp}.

We propose some ideas for future work. First, while making linearly fair features works very well in practice, the fairness could be improved by removing non-linear correlations.  
Second, we can extend our method to more easily address categorical protected variables. In the present method, a categorical variable with alphabet size $n$ becomes a set of $n-1$ bivariate variables. It would be ideal, however, if a method reduced the mutual information between the categorical variable directly, rather than first creating $n-1$ variables, and removed correlations.

\section*{Acknowledgements}
Authors would like to thank Ayush Jaiswal for providing the code for learning adversarial models and feedback on results. Authors also thank Daniel Moyer and Greg Ver Steeg for insightful discussions about the approach. This material is based upon work supported in part by the Defense Advanced Research Projects Agency (DARPA) ) under Contracts No. W911NF-18-C-0011 and HR00111990114. This research is also based upon work supported in part by the Office of the Director of National Intelligence (ODNI), Intelligence Advanced Research Projects Activity (IARPA), via 2017-17071900005. The views and conclusions contained herein are those of the authors and should not be interpreted as necessarily representing the official policies, either expressed or implied, of ODNI, IARPA, or the U.S. Government. The U.S. Government is authorized to reproduce and distribute reprints for governmental purposes notwithstanding any copyright annotation therein.



\section{Appendix}

\subsection{Additional Tables}

We show the comparison of our method with former works on invariant representations in Table \ref{tb:acc_yp_german_adult}. Following the former works, we use accuracy of predicted outcomes (\textit{Acc Y}) and accuracy of protected features (\textit{Acc P}) as performance metrics.

\begin{table}[h!]
    \centering
    \begin{tabular}{lll|ll}
    \toprule
    & \multicolumn{2}{c}{\textit{German}} & \multicolumn{2}{c}{\textit{Adult}}\\
    \cmidrule{2-5}
    \textit{Method} & \textit{Acc Y} & \textit{Acc P} & \textit{Acc Y} & \textit{Acc P}\\
    \hline
    Maj. Class & 0.71  & 0.80 & 0.75 & 0.67 \\ \midrule
    Li \shortcite{Li2014} *& 0.74 &\bf{0.80}& 0.76 &\bf{0.67}\\
    VFAE \shortcite{Louizos2015} * & 0.73 & 0.70 & 0.81  &\bf{0.67}\\
    Xie \shortcite{Xie2017} * & 0.74 &\bf{0.80} & 0.84 &\bf{0.67} \\
    Moyer \shortcite{Moyer2018} * & 0.74 & 0.60 & 0.79 & 0.69 \\
    Jaiswal \shortcite{Jaiswal2018} * &\bf{0.78} & \bf{0.80} & 0.84  &\bf{0.67} \\ \midrule
    Fair Logistic & 0.74 &\bf{0.80} & 0.84 &\bf{0.67} \\
    Fair NuSVM & 0.75 & \bf{0.80} & \bf{0.85} &0.73\\
    Fair AdaBoost & 0.75 & \bf{0.80} & 0.84 & \bf{0.67} \\
    Fair RF & 0.75 & \bf{0.80} & \bf{0.85} & 0.72\\
    Fair MLP & 0.75 & \bf{0.80} & \bf{0.85} & \bf{0.67} \\
    \bottomrule
    \end{tabular}
    \caption{Accuracy of predicted outcomes (\textit{Acc Y}) and protected features (\textit{Acc P}) for the German and Adult datasets. The proposed fair methods (bottom four rows) use $\lambda=0.0$. Higher \textit{Acc Y} indicates better predictions while \textit{Acc P} closer to the majority class baseline indicates fairer predictions. Results marked * were reported by \cite{Jaiswal2019}. Best performance is shown in bold.}
    \label{tb:acc_yp_german_adult}
\end{table}

\subsection{Proof for the Invariant of Parameters in Linear Regression Using Debiased Features}

Consider a linear regression using all the $n_f$ non-protected features $x_i, i=1, \dots, n_f $ and $n_p$ protected features $p_i, i=1, \dots, n_p$.
\begin{equation}
    \hat y = \beta_0 + \sum_{i=1}^{n_f} \beta_i x_i 
                     + \sum_{i=1}^{n_p} \gamma_i p_i.
\end{equation}
Assuming we have created a model using the debiased features $x_i', i=1, \cdots, n_f$, 
\begin{equation}
    \hat y' = \beta_0' + \sum_{i=1}^{n_f} \beta_i' x_i'.
\end{equation}
We now give a mathematical proof that
\begin{equation}
    \beta_i = \beta_i', \forall i=1,\dots,n_f.
    \label{eq:appendix_target}
\end{equation}
For simplicity, we assume that all the features and also the outcome have means equal to $0$ and standard deviations equal to $1$. In this case, the Pearson correlation between features can be calculated as inner products,
\begin{equation}
    \text{Corr}(x,y) = \bm{x} \cdot \bm{y}
\end{equation}
In the equation, $\bm{x}$ and $\bm{y}$ can be non-protected features, protected features or outcome. The bold fond stands for a vector of in $N$-dimension, where $N$ is the number of data points.

without loss of generality, assume that all the protected protected features are orthogonal to each other. (Generally speaking, protected features can be correlated. But we can always find orthogonal basis for them.)
The debiased features $\bm{x}'$  can be calculated in the following way, 
\begin{equation}
    \bm{x}_i' = \bm{x}_i - \sum_{j=1}^{n_p} c_{ij} \bm{p}_j
\end{equation}
, where $c_{ij} = \text{corr}(x_i, p_j) = \bm{x_i}\cdot\bm{p_j}$. Since all features and the outcome has mean equals to $0$, $~\beta_0 = \beta_0' = 0$. Other parameters are solved by the inversion problem below,
\begin{equation}
     \tilde X^T \tilde X \tilde \beta = \tilde X^T y, ~X'^T X' \beta' = X'^T y.
\end{equation}
Let $X = [\bm x_1, \dots, \bm x_{n_f}]$, $P = [\bm p_1, \dots, \bm p_n]$ and $C = [c_{ij}]$. Here $X' = [\bm x_1', \dots, \bm x_{n_f}']$, $\tilde X = [\bm x_1, \dots, \bm x_{n_f}, \bm p_1, \dots, \bm p_{n_p}] = [X~P]$, $\beta = [\beta_1, \dots, \beta_{n_f}]^T$ and finally $\tilde \beta = [\beta_1, \dots, \beta_{n_f}, \gamma_1, \dots, \gamma_{n_p}]^T$. 
 
 Then, we have 
 \begin{equation}
     \tilde X^T \tilde X = 
     \begin{bmatrix} 
        X^T\\
        P^T
    \end{bmatrix}
     \begin{bmatrix} 
        X P
    \end{bmatrix}
    = \begin{bmatrix} 
        X^T X & X^T P \\
        P^T X & P^T P
      \end{bmatrix}.
\end{equation}
Using the assumption that all the protected features are orthogonal to each other and the definition of $c_{ij}$,
\begin{equation}
    \tilde X^T \tilde X = 
      \begin{bmatrix} 
        X^T X & C \\
        C^T   & I
      \end{bmatrix}.
\end{equation}
And we also have 
\begin{equation}
    \tilde X^T y = 
      \begin{bmatrix} 
        X^T y \\
        P^T y
      \end{bmatrix}.
\end{equation}

Now, we consider the inversion problem for regression using debiased features. 
\begin{eqnarray}
    && (X'^T X')_{ij} \\
    & = & \bm x_i' \cdot \bm x_j' \\
    & = & (\bm x_i - \sum_l c_{il} \bm p_l)\cdot(\bm x_j - \sum_k c_{jk} \bm p_k)\\
    &=& \bm x_i \cdot \bm x_j - \bm x_i \cdot \sum_k c_{jk} \bm p_k - \bm x_j \cdot \sum_l c_{il} \bm p_l\\
    && + \sum_l c_{il} c_{jl} \\
    &=& \bm x_i \cdot \bm x_j - \sum_l c_{il} c_{jl}
\end{eqnarray}

\begin{eqnarray}
    (X'^T y)_i &=&  (\bm x_i - \sum_l c_{il} \bm p_l) \cdot \bm y\\
    &=& \bm x_i \cdot \bm y - \sum_l c_{il} \bm p_l \cdot \bm y
\end{eqnarray}

We can see that the rows of matrix $X'^TX'$ and $X'^Ty$ can be obtained by applying the same elementary row reduction steps to $\tilde X^T \tilde X$ and $\tilde X^T y$. To obtain the $i$th row, we perform
\begin{equation}
   \text{Row}_i = \text{Row}_i - \sum_{l=1}^{n_p} c_{il} * \text{Row}_{(i+n_f)}.
\end{equation}
After applying the elementary row reduction steps above to the first $n_f$ rows of the matrix form of $\tilde X^T \tilde X' \tilde \beta = \tilde X^Ty$, we will have
\begin{equation}
    \begin{bmatrix} 
        X'^T X' & 0 \\
        C^T   & I
     \end{bmatrix} \tilde \beta = 
     \begin{bmatrix} 
        X'^T y \\
        P^T y 
     \end{bmatrix}.
\end{equation}
Thus inversion problem $X'^T X' \beta' = X'^T y$ is a sub-problem of $ \tilde X^T \tilde X \tilde \beta = \tilde X^T y$ where the first $n_f$ elements of $\tilde \beta$ gives $\beta'$. Which means that we have proved Eq.(\ref{eq:appendix_target}). It is worth mention that for fairness level $\lambda \neq 0$, the statement of Eq.(\ref{eq:appendix_target}) does not hold.

\end{document}